\documentclass[11pt,a4paper,english]{article}
\pdfoutput=1

\usepackage{lmodern}
\renewcommand{\sfdefault}{lmss}

\usepackage{float}
\usepackage{mathtools}
\usepackage{enumitem}
\usepackage{amsmath}
\usepackage{amsthm}
\usepackage{amssymb}
\usepackage{graphicx}
\usepackage{esint}

\makeatletter

\newif\ifColors
\newif\ifCrop

\newif\ifprstyle

\ifx\affiliation\undefined
  \prstylefalse
\else
  \prstyletrue
  \newif\ifnotoc
  \notoctrue
\fi

\ifprstyle
  \ifColors
    \usepackage[pdfa=true]{hyperref}
  \else
    \usepackage[
      colorlinks=false, pdfa=true, linktocpage=true
    ]{hyperref}
  \fi
\else
  \usepackage{jheppub}
  \newcommand{\email}[1]{\emailAdd{#1}}
\fi

\ifCrop

  \ifx\ifPresentation\undefined
     \usepackage[paperwidth=171.2mm,paperheight=261.2mm]{geometry}
  \else
     \usepackage[paperwidth=330mm,paperheight=233mm]{geometry}
  \fi

  \divide\@tempdima\baselineskip
  \@tempcnta=\@tempdima
  \ifx\ifPresentation\undefined
    \hypersetup{pdfstartview={Fit},pdfpagelayout={TwoPageRight}}
  \else
    \hypersetup{pdfstartview={Fit},pdfpagelayout={SinglePage}}

    \usepackage{everypage}
    \AddEverypageHook{\tikz[remember picture,overlay]{%
      \draw [gray!10] ($(current page.east)+(-130mm,+180mm)$)
         -- ($(current page.east)+(-130mm,-100mm)$);}}
  \fi
\fi

\ifColors
  \hypersetup{
    colorlinks=true,          pdfa=true,
    urlcolor=black!40!blue,   anchorcolor=black!40!blue,
    citecolor=black!40!blue,  filecolor=black!40!blue,
    linkcolor=black!40!blue,  menucolor=black!40!blue,
    pagecolor=black!40!blue,  linktocpage=true
  }
\else
  \hypersetup{
    colorlinks=true,  pdfa=true,
    urlcolor=black,   anchorcolor=black,
    citecolor=black,  filecolor=black,
    linkcolor=black,  menucolor=black,
    pagecolor=black,  linktocpage=true
  }
\fi

\usepackage{bookmark}

\usepackage{amsmath,amsfonts,amssymb,bm,cancel}

\usepackage{color}
\usepackage{colortbl}

\ifprstyle
  
\else
  
\fi

\newcommand{\bSe}{\begin{subequations}}
\newcommand{\eSe}{\end{subequations}}

\newcommand{\bWe}{\begin{widetext}}
\newcommand{\eWe}{\end{widetext}}

\allowdisplaybreaks

\usepackage{tikz}
\usetikzlibrary{shapes,arrows}
\usetikzlibrary{calc}
\usetikzlibrary{positioning}
\usetikzlibrary{decorations.pathreplacing}
\usetikzlibrary{shapes.misc}

\usepackage[scr,scaled=1.1]{rsfso}
\usepackage{eucal}

\DeclareMathAlphabet{\mathsfit}{\encodingdefault}{\sfdefault}{m}{sl}
\SetMathAlphabet{\mathsfit}{bold}{\encodingdefault}{\sfdefault}{bx}{sl}

\ifprstyle\else
\fi

\newcommand{\emitFrontMatter}{
  \ifprstyle
    \begin{abstract}\myAbstract\end{abstract}
  \else
    \ifnotoc
      \abstract{\\[-1.64ex]\hphantom{\hspace{0.28cm}}%
        \parbox{1\columnwidth-0.28cm}%
        {\renewcommand\baselinestretch{1.05}\small\hspace{18mm}\myAbstract}}
    \else
      \abstract{\myAbstract}
    \fi
  \fi

  \keywords{\myKeywords}

  \ifprstyle\else
    \makeatletter
      \def\@fpheader{\myReleaseInfo}
    \makeatother
    \ifnotoc
      \compress
      \renewcommand\afterLogoSpace{}
      \renewcommand\afterSubheaderSpace{}
      \renewcommand\afterProceedingsSpace{}
      \makeatletter
      \renewcommand\ps@titlepage{}
      \makeatother
      \toccontinuoustrue
    \else
      \addtocontents{toc}{\protect\setcounter{tocdepth}{2}}
    \fi
  \fi

  \maketitle

  \ifprstyle\else
    \ifnotoc
      \hrule\bigskip\bigskip
    \else
      \flushbottom
    \fi
  \fi
}

\usepackage{titlesec}

\newcommand{\emitAppendix}{
  \phantomsection
  \addcontentsline{toc}{section}{Appendices}
  \addtocontents{toc}{\protect\setcounter{tocdepth}{2}}

  \makeatletter
   \def\toclevel@section{1}
    \def\toclevel@subsection{2}
  \makeatother

  \addtocontents{toc}{\protect\makeatletter}
  \addtocontents{toc}{\string\let\string\l@chapter\string\l@section}
  \addtocontents{toc}{\string\let\string\l@section\string\l@subsection}
  \addtocontents{toc}{\protect\makeatother}

  \titleformat{\section}{\normalfont\large\bfseries}{Appendix~\thesection~~~}{0em}{}

  \let\oldSection\section
  \renewcommand\section{\bookmarksetupnext{level=2}\oldSection}

  \appendix
}

\pdfpageheight\paperheight
\pdfpagewidth\paperwidth

\providecommand{\tabularnewline}{\\}

\makeatother

\usepackage{babel}
\begin{document}

\ifx \ii \undefined

\renewcommand\tilde[1]{\mkern1mu\widetilde{\mkern-1mu#1}}

\newcommand\NPOne{$N\!+\!1$}

\global\long\def\NPOne{\mbox{\textit{N}+1}}

\global\long\def\ii{\mathrm{i}}
\global\long\def\ee{\mathrm{e}}
\global\long\def\dd{\mathrm{d}}
\global\long\def\ppi{\mathrm{\pi}}
\global\long\def\tr{\mathsf{{\scriptscriptstyle T}}}
\global\long\def\Tr{\operatorname{Tr}}
\global\long\def\op#1{\operatorname{#1}}
\global\long\def\dim{\operatorname{dim}}
\global\long\def\diag{\operatorname{diag}}
\global\long\def\Lie{\mathrm{\mathscr{L}}}

\global\long\def\pp#1{#1^{\prime}}
\global\long\def\matc#1{[#1]}

\global\long\def\mfrac#1#2{\frac{\raisebox{-0.45ex}{\scalebox{0.9}{#1}}}{\raisebox{0.4ex}{\scalebox{0.9}{#2}}}}
\global\long\def\mbinom#1#2{\Big(\begin{array}{c}
 #1\\[-0.75ex]
 #2 
\end{array}\Big)}

\global\long\def\tud#1#2#3{#1{}^{#2}{}_{#3}}
\global\long\def\tdu#1#2#3{#1{}_{#2}{}^{#3}}

\unless\ifColors

\global\long\def\gSector#1{{\color{black}#1}}
\global\long\def\fSector#1{{\color{black}#1}}
\global\long\def\lSector#1{{\color{black}#1}}
\global\long\def\sSector#1{{\color{black}#1}}

\fi\ifColors

\definecolor{red}{rgb}{1,0,0.1}
\definecolor{green}{rgb}{0.0,0.6,0}
\definecolor{blue}{rgb}{0.1,0.1,1}
\definecolor{brown}{rgb}{0.6,0.3,0}
\definecolor{orange}{rgb}{0.8,0.3,0}
\definecolor{magenta}{rgb}{0.9,0.1,1}

\global\long\def\gSector#1{{\color{blue}#1}}
\global\long\def\fSector#1{{\color{red}#1}}
\global\long\def\lSector#1{{\color{green}#1}}
\global\long\def\sSector#1{{\color{magenta}#1}}

\fi

\global\long\def\gMet{\gSector g}
\global\long\def\gLapse{\gSector N}
\global\long\def\gShift{\gSector{\nu}}
\global\long\def\gShiftVec{\gSector{\nu}}
\global\long\def\gESp{\gSector e}
\global\long\def\gSp{\gSector{\hat{g}}}

\global\long\def\gE{\gSector E}

\global\long\def\fMet{\fSector f}
\global\long\def\fLapse{\fSector M}
\global\long\def\fShift{\fSector{\mu}}
\global\long\def\fShiftVec{\fSector{\mu}}
\global\long\def\fESp{\fSector m}
\global\long\def\fSp{\fSector{\hat{f}}}

\global\long\def\fE{\fSector L}
\global\long\def\fEs{\fSector{\hat{L}}}
\global\long\def\fEt{\sSector u}
\global\long\def\fEu{\sSector q}
\global\long\def\fEut{\fSector{u^{0}}}

\global\long\def\sLs{\lSector{\hat{\Lambda}}}
\global\long\def\sLt{\lSector{\lambda}}
\global\long\def\sLtinv{\lSector{\lambda^{-1}}}
\global\long\def\sLv{\lSector v}
\global\long\def\sLp{\lSector p}
\global\long\def\sRs{\lSector{\hat{R}}}
\global\long\def\sRbar{\lSector{\bar{R}}}

\global\long\def\sI{\lSector{\hat{I}}}
\global\long\def\sEta{\lSector{\hat{\delta}}}
\global\long\def\sLP{\lSector{\omega}}

\global\long\def\sDs{\sSector{\hat{D}}}
\global\long\def\sBs{\sSector{\hat{B}}}
\global\long\def\sB{\sSector B}
\global\long\def\delShift{\sSector{\Delta n}}

\fi


\newcommand{\myReleaseInfo}{~\\~}

\newcommand{\myTitle}{Bimetric interactions based on metric congruences}

\newcommand{\myAbstract}{In massive gravity and bigravity, spin-2
interactions are defined in terms of a square root matrix that involves
two metrics. In this work, the interactions are constructed using
a congruence matrix between the metrics. It is established that the
primary square root matrix function is the only power series solution
to the equations of motion for the congruence. Moreover, the shift
vector redefinition that is used in the bimetric ghost-free proofs
follows from the $N+1$ form of the equations of motion. The analysis
also gives an insight into the vielbein formulation of spin-2 interactions
since the bimetric formulation in terms of a congruence is algebraically
equivalent to the unconstrained vielbein formulation.}

\newcommand{\myKeywords}{\bgroup\small Modified gravity, Massive
gravity, Bigravity, Ghost-free bimetric theory\egroup}

\title{\myTitle}

\author{Mikica Kocic}

\affiliation{%
  Department of Physics \& The Oskar Klein Centre,\\
  Stockholm University, AlbaNova University Centre,
  SE-106 91 Stockholm
}

\email{mikica.kocic@fysik.su.se}

\hypersetup{
  pdftitle=\myTitle,
  pdfauthor=Mikica Kocic,
  pdfsubject=Hassan-Rosen ghost-free bimetric theory,
  pdfkeywords={Modified gravity, Massive gravity, Bigravity, %
    Ghost-free bimetric theory}
}

\emitFrontMatter


\section{Introduction}

General relativity is the classical theory of nonlinear self-interactions
for a massless spin-2 field governed by the Einstein\textendash Hilbert
action. The context of this paper are its extensions: de Rham\textendash Gabadadze\textendash Tolley
(dRGT) massive gravity \cite{deRham:2010kj,Hassan:2011hr,Hassan:2011tf}
and the Hassan\textendash Rosen (HR) bimetric theory or bigravity
\cite{Hassan:2011zd,Hassan:2011ea,Hassan:2018mbl}. Massive gravity
is a nonlinear theory of a single massive spin-2 field, while bigravity
is a nonlinear theory of two interacting spin-2 fields having both
massless and massive modes. These theories have been studied extensively
over the past years; for reviews, see \cite{deRham:2014zqa,Hinterbichler:2011tt,Schmidt-May:2015vnx}.

Massive gravity and bigravity are classically consistent theories,
free of instabilities such as the Boulware\textendash Deser ghost
\cite{Boulware:1973my}. This is not a coincidence, but due to the
particular structure of the bimetric scalar potential proposed in
\cite{deRham:2010kj} with a compact general form \cite{Hassan:2011tf},
\begin{equation}
\mathcal{S}^{\mathrm{int}}(\gMet,\fMet)=\sum_{n=0}^{d}\beta_{n}\intop\dd^{d}x\,\sqrt{-\gMet}\,e_{n}\Big(\sqrt{\gMet^{-1}\fMet}\,\Big).\label{eq:bim-pot}
\end{equation}
The potential involves two metric fields $\gMet$ and $\fMet$; it
is specifically constructed in terms of the square root matrix $(\gMet^{-1}\fMet)^{1/2}$
using the elementary symmetric polynomials $e_{n}$ \cite{macdonald:1998a},
where the interaction is parametrized by constants $\beta_{n}$. In
massive gravity, the metric $\gMet$ carries dynamics while $\fMet$
is a reference nondynamical field. In bigravity, both $\gMet$ and
$\fMet$ are dynamical, each having its own Einstein\textendash Hilbert
term. 

The vielbein formulation of spin-2 interactions was constructed in
\cite{Hinterbichler:2012cn}. The vielbein based potential has the
following expression (given in \cite{Hinterbichler:2012cn}, cf.~\cite{Zumino:1970tu}),
\begin{equation}
\mathcal{S}^{\mathrm{int}}(\gMet,\fMet)=\sum_{n=0}^{d}\frac{\beta_{n}}{n!(d\!-\!n)!}\intop\epsilon_{A_{1}A_{2}\cdots A_{d}}\fE^{A_{1}}\wedge\cdots\wedge\fE^{A_{n}}\wedge\gE^{A_{n+1}}\wedge\cdots\wedge\gE^{A_{d}}.\label{eq:viel-pot}
\end{equation}
Here, $\epsilon$ denotes the Levi\textendash Civita symbol, and the
vielbeins are represented by one-forms $\gE^{A}=\tud{\gE}A{\mu}\dd x^{\mu}$
and $\fE^{A}=\tud{\fE}A{\mu}\dd x^{\mu}$. The associated metrics
are,
\begin{equation}
\gMet=\gMet_{\mu\nu}\dd x^{\mu}\dd x^{\nu}=\eta_{AB}\gE^{A}\gE^{B},\qquad\fMet=\fMet_{\mu\nu}\dd x^{\mu}\dd x^{\nu}=\eta_{AB}\fE^{A}\fE^{B},
\end{equation}
where $\eta=\diag(-,+,\cdots,+)$ is the metric of the local Lorentz
frame. In the constrained version, $h=\eta_{AB}\fE^{A}\gE^{B}$ is
symmetric, which is equivalent to having a real square root $(\gMet^{-1}\fMet)^{1/2}=\gE^{-1}\fE$
in matrix notation \cite{Deffayet:2012zc}. Consequently, the constrained
vielbein formulation is equivalent to the metric formulation since
(\ref{eq:bim-pot}) and (\ref{eq:viel-pot}) become equal. The metric
formulation of the multivielbein theory was treated in \cite{Hassan:2012wt}.

\paragraph*{Purpose of this work.}

\begin{figure}[H]
\noindent \begin{centering}
{\small(a)}\includegraphics[width=32mm]{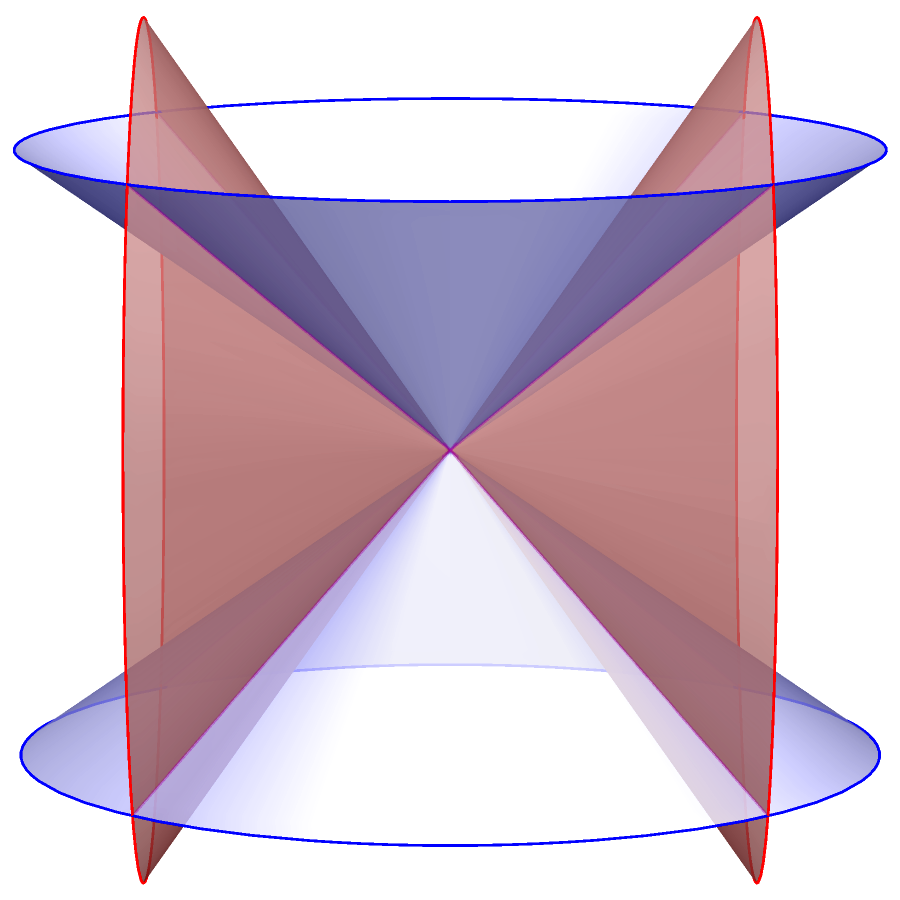}\hspace{18mm}\includegraphics[width=32mm]{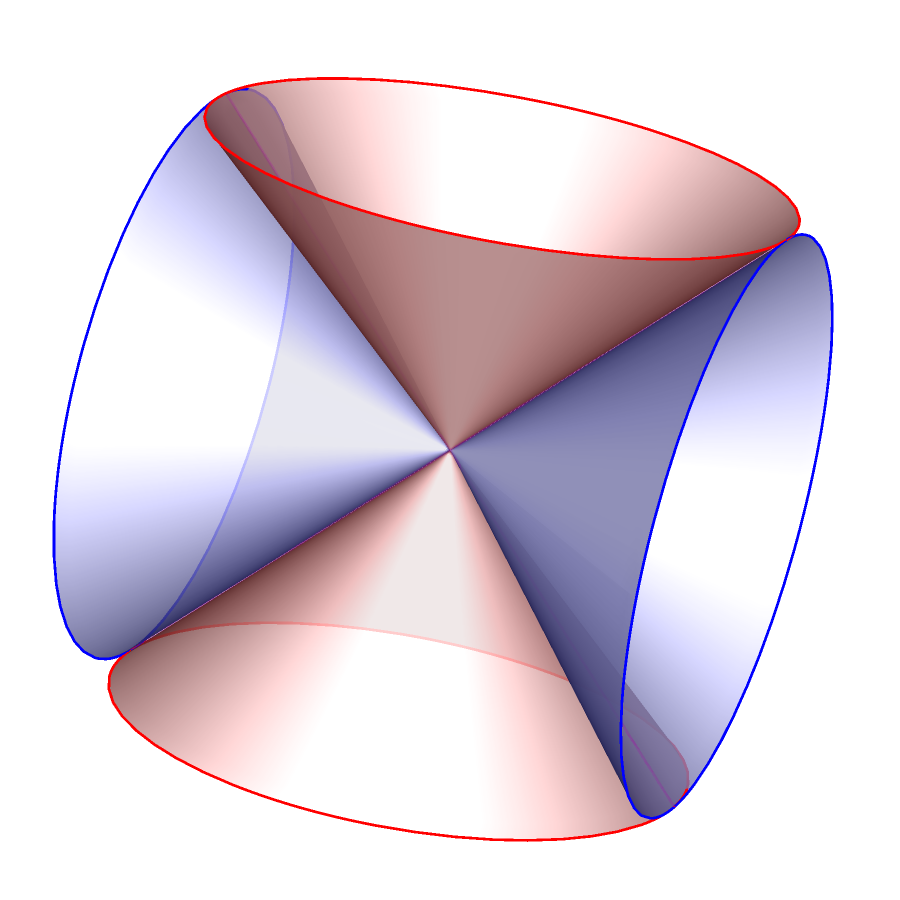}\llap{\small(b)\hspace{28mm}}\hspace{2mm}\vspace{-1ex}
\par\end{centering}
\caption{\label{fig:bad-nc}Null cones for the metric configurations which
do not allow a boosted $\protect\NPOne$ decomposition wherein both
vielbeins are in the triangular form. Configuration (a) does not have
a real square root, and (b) has an infinite number of nonprimary real
square roots that are not tensor fields (for more details about possible
metric configurations see \cite{Hassan:2017ugh}).}
\end{figure}

We address several issues related to bimetric interactions. First,
the question if there exists a more general bimetric formulation of
the potential (\ref{eq:bim-pot}) yet \emph{equivalent} to the vielbein
formulation that is based the symmetric polynomials; then, if it exists,
what governs the selection of the square root in the \emph{constrained}
case? Namely, the square root matrix $(\gMet^{-1}\fMet)^{1/2}$ must
represent a well-defined tensor field. Notwithstanding, it may have
multiple branches, and besides not being real, it can be solved ad
hoc from the matrix equation $\gMet^{-1}\fMet=S^{2}$ which may have
an infinite number of solutions (an example of such $\gMet^{-1}\fMet$
is shown in figure~\ref{fig:bad-nc}b). In the vielbein formulation,
the square root selection is concealed inside the symmetrization condition
giving no definite choice for $S$ \cite{Deffayet:2012zc}.

Another issue concerns the initial value problem for the unconstrained
vielbein formulation. The ghost-free proof in \cite{Hinterbichler:2012cn}
assumes the simultaneous $\NPOne$ decomposition with arbitrarily
boosted vielbeins. This is in general not possible since one cannot
ensure that an arbitrary vielbein $\fE$ can simultaneously be triangularized
with  $\gE$ already being in the triangular form. A typical example
is shown in figure~\ref{fig:bad-nc}a, where the null cones of the
two metrics doubly intersect; in this case the simultaneous $\NPOne$
decomposition of $\gMet$ \& $\fMet$, and the simultaneous triangularization
of $\gE$ \& $\fE$, are not possible.

\paragraph*{Summary of results.}

\begin{figure}[H]
\noindent \begin{centering}
\includegraphics[scale=0.9]{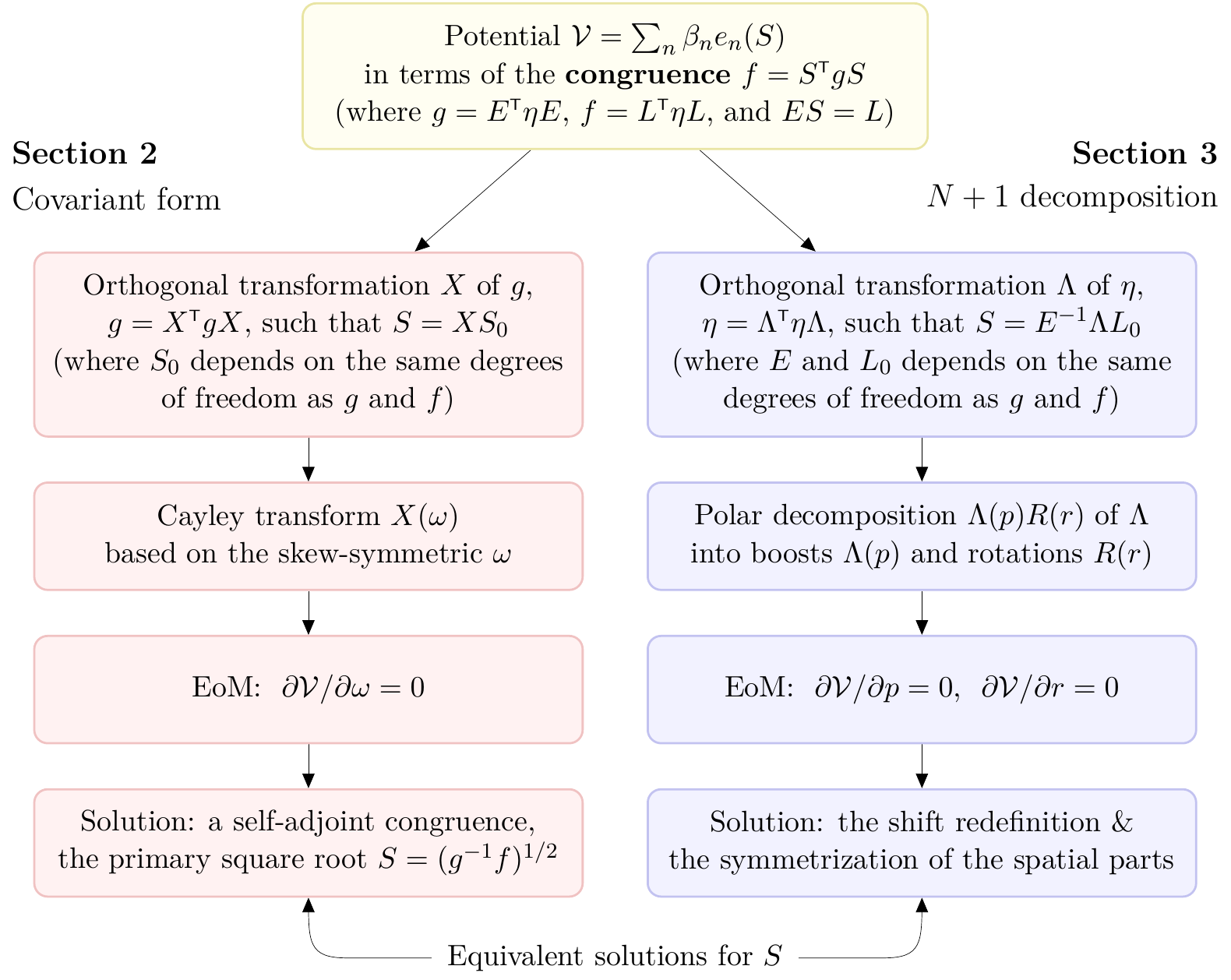}
\par\end{centering}
\caption{\label{fig:sum}Summary of results and paper structure. The starting
point is the bimetric potential whose algebraic structure is based
on the elementary symmetric polynomials $e_{n}(S)$ dependent on a
general congruence $S$ between the metrics $f=S^{\protect\tr}gS$.
The on-shell conditions for $S$ are obtained in two alternative ways:
covariantly in section 2, and using the $N\!+\!1$ decomposition in
section 3. Both approaches give the congruence $S=(g^{-1}f)^{1/2}$.
The derivation in section 2 is more formal than in section 3, and
it shows that the square root is necessarily a primary matrix function.}
\end{figure}

An overview of the paper with the key results is shown in figure~\ref{fig:sum}.
\linebreak{}
In section \ref{sec:cov}, we construct bimetric interactions in terms
of a general congruence $\fMet=S^{\tr}\gMet S$ \linebreak{}
between the metric fields $\gMet$ and $\fMet$. This congruence based
metric formulation is algebraically equivalent to the unconstrained
vielbein formulation~\cite{Hinterbichler:2012cn}. We solve the equation
of motion for the congruence, which necessarily gives the primary
square root $S=(\gMet^{-1}\fMet)^{1/2}$ as the congruence field.
In section~\ref{sec:np1}, the equations are solved using the $\NPOne$
decomposition. The obtained solution is the symmetrization of the
spatial metrics together with the shift vector redefinition that was
used in the bimetric ghost-free proofs~\cite{Hassan:2011tf,Hassan:2011zd}.
The rest of the introduction is devoted to a mathematical background
on metric congruences.\medskip

\smallskip

\paragraph*{Notation.}

The metric signature is mostly positive $(-,+,\cdots,+)$. The spacetime
dimension is $d$. In the space-plus-time decomposition, $d=\NPOne$,
hats over spacetime objects denote their spatial restrictions. The
equations are mostly written in matrix notation. Hence, the expressions
are preferably stated using (1,1)-tensors with the default down-up
contractions. Matrices do not naturally represent metrics and require
transposes on the left side of the metric symbol in matrix notation.
Examples of how to restore the indices are given in appendix \ref{app:indices}.
Spacetime (world) indices are denoted by $\mu,\nu,\ldots$ and their
spatial restrictions by $i,j,\ldots$, while Lorentz (local frame)
indices are denoted by $A,B,\ldots$ and their spatial restrictions
by $a,b,\ldots$.

\subsection{Metric congruences}

Here we review some basic properties of symmetric bilinear forms (metrics)
and their isometries (congruent transformations or congruences); for
more details see \cite{Scharlau:1985,Lang:2002,Meinrenken:2013}.
Two examples of congruences are orthogonal transformations and moving
frames (vielbeins).

Let $\gMet$ be a nondegenerate symmetric bilinear form on a finite-dimensional
real vector space $V$. The pair $(V,\gMet)$ is called a real symmetric
bilinear space. Two symmetric bilinear spaces $(V,\gMet)$ and $(\tilde{V},\fMet)$
are isometric iff there exists an invertible linear transformation
$S:\tilde{V}\to V$ such that,
\begin{equation}
\fMet(v,w)=\gMet\bigl(S(v),S(w)\bigr),\qquad\text{for all }v,w\in\tilde{V}.
\end{equation}
That is, $\fMet$ is a precomposition of $\gMet$ with the map $S$
(or the pullback of $\gMet$ by $S$). This corresponds to the matrix
congruence $\fMet=S^{\tr}\gMet S$ in some basis. In linear algebra,
the linear transformation $S$ is usually referred to as an isometry.
To avoid possible confusion with the Killing symmetries of the metric
fields, we adopt another frequently used name ``congruence'' or
``congruent transformation.'' By Sylvester's law of inertia, the
congruent transformations preserve the signature of the symmetric
bilinear forms. They form a group of automorphisms on the same vector
space $\tilde{V}=V$.

The point-wise notion of a congruence can be lifted from linear algebra
to differential geometry (i.e., from a tensor to a tensor field).
A possible obstruction is that the congruence field might not be defined
as a global section of the fiber bundle. A typical example is a vielbein,
which globally exists iff the manifold is parallelizable \cite{Hawking:1973large}.
Nevertheless, one can always assume the local existence of sections,
which we employ here.

\paragraph*{Adjoint maps.}

Two linear transformations $A$ and $A^{\prime}$ are adjoint with
respect to $\gMet$, iff 
\begin{equation}
\gMet\bigl(A(v),w\bigr)=\gMet\bigl(v,A^{\prime}(w)\bigr),\qquad\text{for all }v,w\in V.
\end{equation}
In matrix notation we have $A^{\prime}=\gMet^{-1}A^{\tr}\gMet$. This
relation can be used to define the adjoint of~$A$ (having a regular
$\gMet$). A self-adjoint transformation $A$ is such that $A^{\prime}\equiv A$.
An example of self-adjoint transformations with respect to $\gMet$
are the matrix functions of $\gMet^{-1}\fMet$. 

\paragraph*{Orthogonal group.}

An orthogonal transformation is a congruence $X$ whose adjoint is
equal its inverse, $X^{\prime}\equiv X^{-1}$. Beware that ``adjointness''
is always stated with respect to some symmetric bilinear form (here
$\gMet$). The orthogonal group $\op O(V,\gMet)$ comprise all the
automorphisms which preserve $\gMet$, that is, $\gMet=X^{\tr}\gMet X$.
Note that a congruence between $(\tilde{V},\fMet)$ and $(V,\gMet)$
always contains excessive degrees of freedom because it can only be
determined up to a residual orthogonal transformation of either $\gMet$
or $\fMet$.

\subsection{Parametrization of orthogonal transformations}

A special orthogonal transformation $X$ can be parametrized by a
skew-symmetric bilinear form $\sLP=-\sLP^{\tr}$ using the Cayley
transform \cite{Golub:1996},
\begin{equation}
X\coloneqq(\gMet+\sLP)^{-1}(\gMet-\sLP).\label{eq:XCayley}
\end{equation}
It is straightforward to verify that $\gMet=X^{\tr}\gMet X=\gMet^{\tr}$
since,
\begin{equation}
\sLP=X^{\tr}\sLP X=-\sLP^{\tr},\qquad\gMet+\sLP=X^{\tr}(\gMet+\sLP)X.
\end{equation}
The parametrization (\ref{eq:XCayley}) is special since $\det X=1$.
In $d$-dimensions, $X$ contains $d(d\!-\!1)/2$ degrees of freedom
in the components of~$\sLP$. Any $X$ can be factored as a noncommutative
product of orthogonal transformations $X=X_{1}X_{2}\cdots$ where
each $X_{i}$ contains partial degrees of freedom in the corresponding
skew-symmetric $\sLP_{i}$. 

\paragraph*{The $\protect\NPOne$ parametrization of boosts and rotations. }

Here we present a recursive definition of the orthogonal transformation,
which is useful to triangularize vielbeins in their entire form. The
same procedure is implicitly employed in the Cholesky decomposition
to factor positive definite matrices into the product of a lower triangular
matrix and its transpose~\cite{Golub:1996}; see also the Cholesky\textendash Banachiewicz
and Cholesky\textendash Crout algorithms \cite{Obsieger:2015nm2}.

Let $\eta$ be a $d$-dimensional metric of either Lorentzian or Euclidean
signature, where $\sEta$ is its $(d\!-\!1)$-dimensional restriction,
\begin{equation}
\eta=\diag(\sigma,1,...,1),\qquad\sigma=\pm1,\qquad\sEta=\diag(1,...,1).
\end{equation}
The orthogonal transformation $\eta=X^{\tr}\eta X$ can be parametrized
\cite{Meinrenken:2013},
\begin{gather}
X\coloneqq\begin{pmatrix}\varepsilon x &  & \varepsilon\sigma\xi^{\prime}\\
\xi &  & \hat{X}
\end{pmatrix}\!\begin{pmatrix}1 &  & 0\\
0 &  & \sRs
\end{pmatrix}\!,\quad x\coloneqq\sqrt{1-\sigma\xi^{\prime}\xi},\quad\hat{X}\coloneqq\sqrt{\sI-\sigma\xi\xi^{\prime}},\quad\xi^{\prime}\coloneqq\xi^{\tr}\sEta,\label{eq:Xdec}
\end{gather}
where $\xi$ is a $(d\!-\!1)$-dimensional vector, $\sEta=\sRs^{\tr}\sEta\sRs$
is an orthogonal transformation of the Euclidean restriction, $\sI$
is the identity map, and $\varepsilon=\pm1$ denotes an optional reflection
(ignored in the following). The parameter space is confined to $\sigma\xi^{\prime}\xi\le1$,
which is satisfied for any $\xi$ in the Lorentzian case $\sigma=-1$.

The factorization (\ref{eq:Xdec}) is recursive. For $d=4$ and $\sigma=-1$,
we have $X=X_{1}X_{2}X_{3}$ where $X_{1}$ comprises three-parameter
boosts of the 4$\times$4 Minkowski metric, $X_{2}$ contains two-parameter
rotations of the 3$\times$3 Euclidean metric, and $X_{3}$ is a one-parameter
rotation of the 2$\times$2 Euclidean metric. This effectively gives
a polar decomposition of the orthogonal transformation $X$ where
the Lorentz boosts $X_{1}$ are parametrized by an arbitrary spatial
Lorentz vector $\sLp$ (see Proposition 1.13 in \cite{Meinrenken:2013}),
\begin{gather}
X_{1}=\Lambda=\begin{pmatrix}\sLt &  & \sLp^{\prime}\\
\sLp &  & \sLs
\end{pmatrix}\!,\quad\sLt\coloneqq\sqrt{1+\sLp^{\prime}\sLp},\quad\sLs\coloneqq\sqrt{\sI+\sLp\sLp^{\prime}},\quad\sLp^{\prime}\coloneqq\sLp^{\tr}\sEta.\label{eq:boost}
\end{gather}
The nonvanishing components of $\sLP$ in the corresponding Cayley
parametrization (\ref{eq:XCayley}) are $\sLP_{0a}=\sLp^{b}\delta_{ba}/(\sLt+1)$.
Also note that,
\begin{equation}
\sLs=\sqrt{\sI+\sLp\sLp^{\prime}}=\sI+\frac{1}{\sLt+1}\sLp\sLp^{\prime},\qquad\sLs^{-1}=\sI-\frac{1}{\sLt(\sLt+1)}\sLp\sLp^{\prime}.
\end{equation}
The boosts can be reparametrized through $\sLv\coloneqq\sLp/\sLt$
where,
\begin{gather}
\sLp=\sLs\sLv=\sLt\sLv,\qquad\sLp^{\prime}=\sLv^{\tr}\sLs^{\tr}\sEta=\sLt\sLv^{\tr}\sEta,\qquad\sLv^{\tr}\sEta\sLv<1.
\end{gather}
Similar expressions hold for rotations. For instance, $X_{3}$ reads,
\begin{equation}
X_{3}=\begin{pmatrix}\sqrt{1-\varrho^{2}} &  & -\varrho\\
\varrho &  & \sqrt{1-\varrho^{2}}
\end{pmatrix}\!,\qquad\varrho^{2}\le1.
\end{equation}

\medskip

\section{Congruence based bimetric scalar potential}

\label{sec:cov}

We consider the scalar potential based on the elementary symmetric
polynomials $e_{n}$,
\begin{equation}
\mathcal{S}^{\mathrm{int}}(\gMet,\fMet)=\intop\dd^{d}x\,\sqrt{-\gMet}\,\mathcal{V}(S),\qquad\mathcal{V}(S)\coloneqq\sum_{n=0}^{d}\beta_{n}e_{n}(S),\label{eq:Vpot}
\end{equation}
where $S$ is an arbitrary congruence between the metric fields $\gMet$
and $\fMet$,\footnote{~Note that (\ref{eq:cong}) is not the most general congruence between
$\gMet$ and $\fMet$ since the metrics can be on different manifolds.
Then, to write (\ref{eq:cong}), we need a diffeomorphism between
the manifolds (with the pullback of one of the metrics), which in
turn introduces a diagonal group of common diffeomorphisms. This generalization,
however, does not affect the presented analysis (see appendix \ref{app:bimpot}
for more details).}
\begin{equation}
\fMet=S^{\tr}\gMet S,\qquad\fMet_{\mu\nu}=\tud S{\rho}{\mu}\tud S{\sigma}{\nu}\gMet_{\rho\sigma}.\label{eq:cong}
\end{equation}
The congruence $S$ is determined up to a local orthogonal transformation
$X$ of $\gMet$,
\begin{equation}
\gMet=X^{\tr}\gMet X,
\end{equation}
such that $S=XS_{0}$, where $S_{0}$ possibly depends on the degrees
of freedom in $\gMet$ and $\fMet$, but not on $X$. The additional
components of the congruence are removed by on-shell conditions for
$X$. All the fields are assumed to be regular since any singularity
punctures the manifold. 

The orthogonal transformation $X$ can be parametrized by a skew-symmetric
tensor field $\sLP$ using the Cayley transform (\ref{eq:XCayley}).
The Einstein\textendash Hilbert term which involves $\gMet$ is not
affected by $X$. Hence, the equations of motion for $S$ are obtained
by varying the potential $\mathcal{V}$, which we summarize in the
following.\footnote{~A similar variation was done in \cite{Hassan:2012wt}; therein,
however, $S$ was a priori assumed to be the square root.} The full derivation is in appendix~\ref{app:deriv}. 

The variation of $\mathcal{V}$ with respect to $X$ reads,
\begin{align}
\delta\mathcal{V} & =\sum_{n=1}^{d}\beta_{n}\sum_{k=1}^{n}(-1)^{k}e_{n-k}(S)\,\Tr\big(X^{-1}S^{k}\delta X\big),
\end{align}
where the further variation of $X$ with respect to $\sLP$ gives,
\begin{gather}
\delta X=-2(\gMet+\sLP)^{-1}\delta\sLP\,(\gMet+\sLP)^{-1}\gMet.
\end{gather}
Substituting $\delta X$ into $\delta\mathcal{V}$ yields,
\begin{align}
\delta\mathcal{V} & =-2\sum_{n=1}^{d}\beta_{n}\sum_{k=1}^{n}(-1)^{k}e_{n-k}(S)\,\Tr\left[(\gMet-\sLP)^{-1}\gMet S^{k}(\gMet+\sLP)^{-1}\delta\sLP\right]\!,
\end{align}
For the skew-symmetric $\sLP$, we have $\delta\sLP^{\tr}=-\delta\sLP$
and it holds, 
\begin{equation}
\partial\Tr(C\sLP)/\partial\sLP=(C-C^{\tr})/2,
\end{equation}
where $C$ is an arbitrary (2,0)-tensor. Hence (see appendix \ref{app:deriv}),
\begin{align}
\frac{\partial\mathcal{V}}{\partial\sLP} & =-\sum_{n=1}^{d}\beta_{n}\sum_{k=1}^{n}(-1)^{k}e_{n-k}(S)\left\{ (\gMet-\sLP)^{-1}\left[\gMet S^{k}-(\gMet S^{k})^{\tr}\right](\gMet+\sLP)^{-1}\right\} \!.
\end{align}
Having the nonsingular $\gMet$ (and $\gMet\pm\sLP$), the equations
of motion $\partial\mathcal{V}/\partial\sLP=0$ become,
\begin{align}
\sum_{n=1}^{d}\beta_{n}\sum_{k=1}^{n}(-1)^{k}e_{n-k}(S)\left[S^{k}-(\gMet^{-1}S^{\tr}\gMet){}^{k}\right] & =0.\label{eq:X-eom1}
\end{align}
The self-adjoint $S=\gMet^{-1}S^{\tr}\gMet$ trivially solves (\ref{eq:X-eom1}),
and since the $\beta$-parameters are arbitrary, this choice is unique
for the independent $\gMet$ and $\fMet$.  When combined with $\fMet=S^{\tr}\gMet S$,
the self-adjoint $S$ yields the equation $S^{2}=\gMet^{-1}\fMet$,
which is solved by a matrix function $S=\sqrt{\gMet^{-1}\fMet}$. 

In the following we shall investigate in more detail the structure
of all possible solutions to~(\ref{eq:X-eom1}), showing that the
primary square root is indeed the unique choice for the congruence
field. Let us express~(\ref{eq:X-eom1}) in terms of,
\begin{equation}
A\coloneqq\gMet^{-1}\fMet.
\end{equation}
Since $\fMet=S^{\tr}\gMet S$ implies $\gMet^{-1}S^{\tr}\gMet=AS^{-1}$,
we have,
\begin{equation}
\sum_{n=1}^{d}\beta_{n}\sum_{k=1}^{n}(-1)^{k}e_{n-k}(S)\left[S^{k}-(AS^{-1})^{k}\right]=0.\label{eq:X-eom}
\end{equation}
This is a nonlinear matrix equation with respect to $S$, depending
only on the nonsingular~$A$. Beside the powers of $S$, the equation
contains the elementary symmetric polynomials of~$S$. The elementary
symmetric polynomials are the principal scalar invariants, which are
part of the Cayley\textendash Hamilton theorem. Subsequently, any
analytic (power series) solution to the equation (\ref{eq:X-eom})
is always a point-wise polynomial. Therefore, (\ref{eq:X-eom}) possibly
has three kinds of solutions:
\begin{enumerate}[parsep=0pt,leftmargin=5em,label=(\roman*)]
\item $S=F(A)$ as a primary matrix function of $A$,
\item $S=F(A)$ as a nonprimary\emph{ }matrix function of $A$, 
\item $S$ is an isolated (incident) solution that is not a function of
$A$.
\end{enumerate}
In the first two cases, the solution is a matrix \emph{function}.
The matrix functions can be defined in many equivalent ways: by Jordan
canonical form, polynomial interpolation, and Cauchy integral theorem
\cite{Horn:1994}. All these definitions produce \emph{primary} matrix
functions. 

A \emph{nonprimary} matrix function is an ``equation solving function''
which cannot be expressed as a primary matrix function \cite{Horn:1994,Higham:2008}.
An example of a nonprimary function is the square root of a matrix
having the same eigenvalue $a$ in different Jordan blocks where the
chosen signs are not the same in $\pm\sqrt{a}$ . Only if $F$ is
a primary function, $F(A)$ is a polynomial in $A$ for all $A$.
Nonprimary matrix functions do not allow perturbations \cite{Konstantinov:2003pt}.

Now, for any symmetric and nonsingular $\gMet$ and $\fMet$, Corollary
1.34 from \cite{Higham:2008} asserts that $\gMet\,F(\gMet^{-1}\fMet)$
and $\fMet\,F(\gMet^{-1}\fMet)$ are also symmetric. This holds regardless
of $F$ being primary or nonprimary. Therefore, we necessarily have
$\gMet S=S^{\tr}\gMet$ and,
\begin{equation}
A=\gMet^{-1}\fMet=S^{2},\quad F(A)=\sqrt{A}.
\end{equation}
However, $S$ is a well-defined tensor field only if $F(A)$ is a
primary function ($S$ can only then be expressed as a polynomial
in $A$). This governs an unambiguous definition of the bimetric theory
given in \cite{Hassan:2017ugh}, which warrants the existence of a
spacetime interpretation by singling out the principal square root
whose eigenvalues lie in the open right complex half-plane, in which
case the square root is unique. 

\begin{figure}
\noindent \begin{centering}
\hspace{-7mm}\includegraphics[scale=0.88]{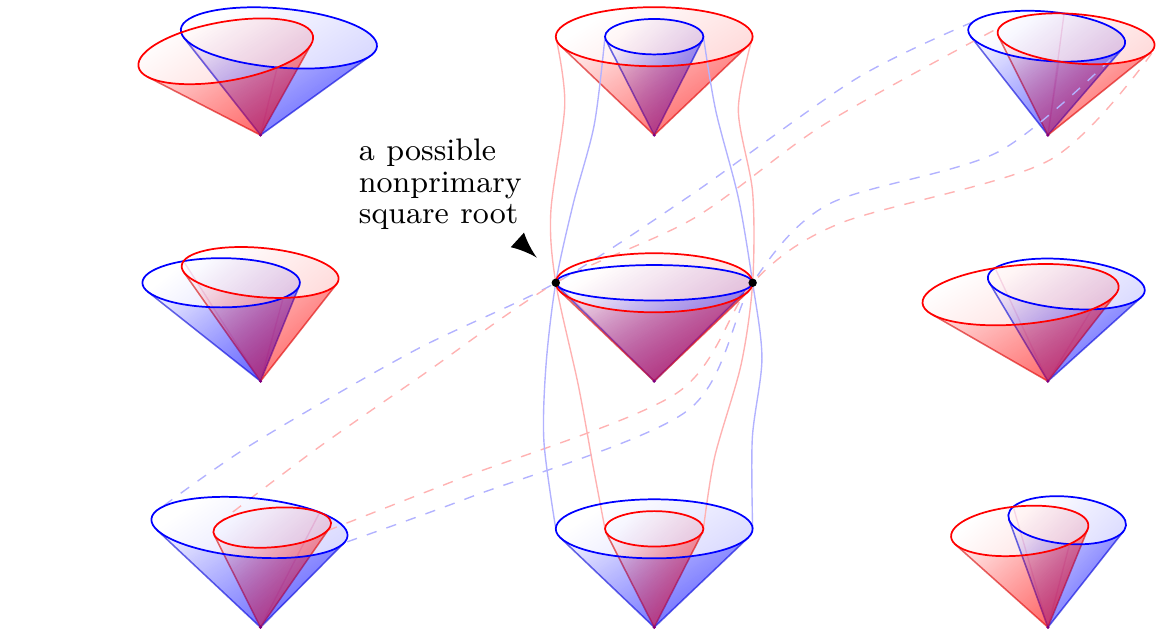}\vspace{-1ex}
\par\end{centering}
\caption{\label{fig:ncfield}The field of primary square roots where one point
(or maybe some region) admits an infinite number of nonprimary solutions
in addition to the primary ones. A well-defined interaction requires
that different paths always agree on the chosen branch of the square
root at the crossings.}
\end{figure}

An example of how a nonprimary solution can be encountered in the
field of primary square roots is shown in figure \ref{fig:ncfield}.
This happens whenever $\gMet^{-1}\fMet$ has the same eigenvalue in
different Jordan blocks. A peculiar metric configuration that only
has nonprimary real square roots (with no real primary roots) is shown
in figure \ref{fig:bad-nc}b.

As earlier noted, the orthogonal transformation $X$ can be factored
into several pieces by splitting the degrees of freedom, $X(\sLP)=X(\sLP_{1})X(\sLP_{2})\cdots X(\sLP_{n})$.
This will result in the equations of motion that form a \emph{coupled
system} for the parameters $\sLP_{1}$, $\sLP_{2}$, ..., $\sLP_{n}$.
Such a case emerges in the following section where we employ the $\NPOne$
decomposition.

The presented analysis is covariant; no particular space-plus-time
decomposition was used or assumed. Nevertheless, the on-shell congruence
$S$ (the \emph{real} primary square root of $\gMet^{-1}\fMet$) enables
the foliation of spacetime with common spacelike hypersurfaces \cite{Hassan:2017ugh}.
The proper $\NPOne$ spacetime foliation is a prerequisite for the
ghost-free proofs that are based on the canonical formalism in the
metric \cite{Hassan:2011tf,Hassan:2011zd,Hassan:2011ea,Hassan:2018mbl}
and the vielbein formulation \cite{Hinterbichler:2012cn}.

\section{Congruences in the $\protect\NPOne$ formalism}

\label{sec:np1}

In this section we derive the equations of motion for the congruence-based
potential in the $\NPOne$ formalism. We shall see that the variation
of this kind of potential gives the shift redefinition from \cite{Hassan:2011tf,Hassan:2011zd}.
The results are also applicable to the equations of motion for the
boost parameter $\sLp$ of the vielbein based potential from \cite{Hinterbichler:2012cn}
when given in the $\NPOne$ form.

\subsection{The $\protect\NPOne$ decomposition}

We first recall how the space-plus-time split \cite{Gourgoulhon:2012trip,Arnowitt:1962hi,York:1979aa}
works for metrics and vielbeins. One can always find a coordinate
patch where one of the metrics is properly $\NPOne$ decomposed, for
instance $\gMet$,
\begin{equation}
\gMet=\begin{pmatrix}-\gLapse^{2}+\gShiftVec^{\tr}\gSp\gShiftVec &  & \gShiftVec^{\tr}\gSp\\
\gSp\gShiftVec &  & \gSp
\end{pmatrix}\!.
\end{equation}
Here, $\gLapse$ is the lapse function, $\gShiftVec$ is the shift
vector, and $\gSp$ is the spatial projection of $\gMet$. The lapse
and the shift are already parts of the timelike vector in a vielbein.
The spatial metric can further be factored $\gSp=\gESp^{\tr}\sEta\gESp$,
which fully defines $\gMet$ in terms of the vielbein $\gE$,\footnote{~If we do not decompose $\gSp$, we end up with the $D$ and $Q$
variables introduced in \cite{Hassan:2011tf}; this favors one metric
(more precise, the metric $\fMet$), and the expressions become ``asymmetric.''
Consequently, the duality between the metrics on the exchange $\gMet\leftrightarrow\fMet$
and $\beta_{n}\leftrightarrow\beta_{d-n}$ would not be explicit.}
\begin{equation}
\gMet=\gE^{\tr}\eta\gE,\qquad\gE=\begin{pmatrix}\gLapse &  & 0\\
\gESp\,\gShiftVec &  & \gESp
\end{pmatrix}\!,\qquad\gE^{-1}=\begin{pmatrix}1/\gLapse &  & 0\\
-\gShiftVec/\gLapse &  & \gESp^{-1}
\end{pmatrix}\!.\label{eq:gE}
\end{equation}
On the other hand, the simultaneous $\NPOne$ split of $\fMet$ is
not possible in general. Nevertheless, $\fMet$ can always be decomposed
in an arbitrary non-null chart into the time restriction $\fMet^{00}$
of $\fMet^{-1}$, the nonsingular space restriction $\fSp$ of $\fMet$
(not necessarily positive definite), and the remaining shear $\fShiftVec$
(an apparent ``shift'' vector),
\begin{equation}
\fMet=\begin{pmatrix}(\fMet^{00})^{-1}+\fShiftVec^{\tr}\fSp\fShiftVec &  & \fShiftVec^{\tr}\fSp\\
\fSp\fShiftVec &  & \fSp
\end{pmatrix}\!,\qquad\fMet^{-1}=\begin{pmatrix}\fMet^{00} &  & -\fMet^{00}\fShiftVec^{\tr}\\
-\fMet^{00}\fShiftVec &  & \fSp^{-1}+\fMet^{00}\fShiftVec\fShiftVec^{\tr}
\end{pmatrix}\!.\label{eq:fMet}
\end{equation}
The sign of $\fMet^{00}$ is arbitrary and depends on the chosen spacetime
foliation, where $\fMet^{00}$ is negative if and only if $\fSp$
is positive definite. In the same chart, $\fMet$ can be given in
terms of a general vielbein $\fE$, 
\begin{equation}
\fMet=\fE^{\tr}\eta\fE,\qquad\fE=\begin{pmatrix}\fEut+\fEt^{\tr}\fEu &  & \fEt^{\tr}\\
\fEs\fEu &  & \fEs
\end{pmatrix}\!,\label{eq:L-2}
\end{equation}
where $\fEs$, $\fEu$, $\fEt$, and $\fEut$ are arbitrary, and $\fEs$
is nonsingular. Equating (\ref{eq:L-2}) and (\ref{eq:fMet}) yields,
\begin{equation}
\fSp=-\fEt\fEt^{\tr}+\fEs^{\tr}\sEta\fEs,\qquad\fShiftVec=\fEu-\fEut\fSp^{-1}\fEt.\label{eq:fLLsp}
\end{equation}
Only in the case of the proper space-plus-time foliation, we have
a real lapse function $\fLapse$,
\begin{equation}
\fLapse^{2}=-(\fMet^{00})^{-1}>0,
\end{equation}
and the vielbein $\fE$ can be put into a triangular form.

\paragraph*{Boosted vielbeins. }

The formal claim is that a general vielbein can be triangularized
by a local Lorentz transformation of the form (\ref{eq:boost}) if
and only if the apparent lapse of the associated metric is real in
a given chart \cite{Kocic:2018ddp}. In other words, the condition
on the coordinate system to be able to extract a real $\fLapse$ from
(\ref{eq:fMet}) is the same as to put $\fE$ into the triangular
form $\fE_{0}$ by $\Lambda^{-1}$ so the triangular $\fE_{0}$ is
boosted to $\fE=\Lambda\fE_{0}$,
\begin{equation}
\Lambda=\begin{pmatrix}\sLt &  & \sLv^{\tr}\sEta\sLs\\
\sLs\sLv &  & \sLs
\end{pmatrix}\!,\quad\fE_{0}=\begin{pmatrix}\fLapse &  & 0\\
\fESp\fShiftVec &  & \fESp
\end{pmatrix}\!,\quad\fE=\begin{pmatrix}\sLt^{-1}\fLapse+\sLs\fESp\,(\fLapse\fESp^{-1}\sLv+\fShiftVec) &  & \sLv^{\tr}\sEta\sLs\fESp\\
\sLs\fESp\,(\fLapse\fESp^{-1}\sLv+\fShiftVec) &  & \sLs\fESp
\end{pmatrix}\!.\label{eq:L-1}
\end{equation}
Comparing (\ref{eq:L-1}) and (\ref{eq:L-2}), one concludes that
only those general vielbeins that satisfy,
\begin{equation}
\fEt^{\tr}(\fEs^{\tr}\sEta\fEs)^{-1}\fEt<1,\label{eq:L-3}
\end{equation}
can be triangularized by a Lorentz transformation since the boosts
are restricted to an open ball $\sLv^{\tr}\sEta\sLv<1$. This is the
same condition as for $\fMet^{00}<0$, that is, $\fSp$ to be positive
definite $\fSp>0$ which follows from (\ref{eq:fLLsp}). For more
details see Lemma 2 in \cite{Kocic:2018ddp}.

\subsection{Variation of the $\protect\NPOne$ form of the potential}

We again start from the bimetric potential (\ref{eq:Vpot}) given
in terms of a congruence $\fMet=S^{\tr}\gMet S$, where $S$ is now
related to the vielbeins (\ref{eq:gE}) and (\ref{eq:L-2}) by $S=\gE^{-1}\fE$.
This form of the potential is algebraically equivalent to the unconstrained
vielbein formulation \cite{Hinterbichler:2012cn}.

The $\NPOne$ form of the potential reads (see appendix \ref{app:deriv}
for the derivation),
\begin{equation}
\gLapse e_{n}(S)=\gLapse e_{n}(\sBs)+\fEut e_{n-1}(\sBs)+\sum_{k=1}^{n}(-1)^{k-1}e_{n-k}(\sBs)\,\fEt^{\tr}\sBs^{k-1}(\fEu-\gShiftVec),
\end{equation}
where $\sBs\coloneqq\gESp^{-1}\fEs$ and $e_{n}(\sBs)=0$ for $n>d-1$.
The potential is manifestly linear in $\gLapse$, $\gShiftVec$, $\fEut$,
and $\fEu$. Note that $\sBs$ is a congruence,
\begin{gather}
\sBs^{\tr}\gSp\sBs=\fEs^{\tr}\sEta\fEs=\fSp+\fEt\fEt^{\tr}.
\end{gather}
Let us consider a coordinate patch where both $\fMet$ and $\gMet$
admit the proper $\NPOne$ decomposition, that is, a patch where both
$\gE$ and $\fE$ can be simultaneously triangularized. In this case,
we have from (\ref{eq:L-1}) and (\ref{eq:L-2}),\bSe
\begin{alignat}{2}
\fEut & =\sLt^{-1}\fLapse, & \qquad & \text{boosted lapse},\\
\fEu & =\fLapse\fESp^{-1}\sLv+\fShiftVec, &  & \text{geometric mean shift},\\
\fEt^{\tr} & =\sLv^{\tr}\sEta\sLs\fESp=\sLv^{\tr}\sEta\fEs, &  & \text{space/time boost},\\
\fEs & =\sLs\fESp, &  & \text{boosted spatial vielbein}.
\end{alignat}
\eSe Hence, $\sBs=\gESp^{-1}\sLs\fESp$ and,
\begin{align}
\gLapse e_{n}(S) & =\gLapse e_{n}(\sBs)+\sLt^{-1}\fLapse e_{n-1}(\sBs)\,+\nonumber \\
 & \qquad+\,\sum_{k=1}^{n}(-1)^{k-1}e_{n-k}(\sBs)\,\sLv^{\tr}\sEta\fEs\sBs^{k-1}(\fLapse\fESp^{-1}\sLv+\fShiftVec-\gShiftVec).\label{eq:NenS1}
\end{align}
This equation is linear in $\gLapse$, $\gShiftVec$, $\fLapse$,
and $\fShiftVec$, which is a necessary condition for the ghost-free
proofs \cite{Hassan:2011tf,Hassan:2011zd,Hassan:2011ea,Hassan:2018mbl,Hinterbichler:2012cn}.
Note, however, that we still do not have a square root at this point;
the velocity vector $\sLv$ and the residual spatial rotation of either
$\gESp$ or $\fESp$ are not constrained.

To vary (\ref{eq:NenS1}) with respect to $\sLv$, the potential must
be rewritten so the elementary symmetric polynomials conveniently
depend only on the spatial vielbeins $\gESp$ and $\fESp$. The derivation
is lengthy and relegated to an ancillary Mathematica notebook (wherein
the calculations are also verified). A final form that is suitable
for variation reads,
\begin{gather}
\gLapse\,e_{n}(S)=\gLapse\biggl\{ e_{n}(\fESp\gESp^{-1})+\sum_{k=1}^{n}(-1)^{k-1}e_{n-k}(\fESp\gESp^{-1})\sLp^{\tr}\sEta\,(\fESp\gESp^{-1})^{k}\,\gESp\,\Bigl[\frac{1}{\sLt+1}\gESp^{-1}\sLp-\gLapse^{-1}\gShiftVec\Bigr]\biggr\}\nonumber \\
\quad+\,\fLapse\biggl\{ e_{d-n}(\gESp\fESp^{-1})+\sum_{k=1}^{d-n}(-1)^{k-1}e_{d-n-k}(\gESp\fESp^{-1})\sLp^{\tr}\sEta\,(\gESp\fESp^{-1})^{k}\,\fESp\,\Bigl[\frac{1}{\sLt+1}\fESp^{-1}\sLp+\fLapse^{-1}\fShiftVec\Bigr]\biggr\}.\label{eq:NenS2}
\end{gather}
Introducing the derivatives of the elementary symmetric polynomials,
\begin{equation}
Y_{n}(X)\coloneqq\sum_{k=0}^{n}(-1)^{n+k}e_{k}(X)\,X^{n-k}=\partial e_{n+1}(X)/\partial X^{\tr},
\end{equation}
the equation (\ref{eq:NenS2}) can be written,
\begin{align}
\gLapse e_{n}(S) & =\gLapse e_{n}\bigl(\fESp\gESp^{-1}\bigr)+\fLapse e_{n-1}\bigl(\fESp\gESp^{-1}\bigr)\nonumber \\
 & \qquad+\,\sLp^{\tr}\sEta\,Y_{n-1}\bigl(\fESp\gESp^{-1}\bigr)\,\fESp\,\Bigl[\frac{1}{\sLt+1}(\gLapse\gESp^{-1}+\fLapse\fESp^{-1})\,\sLp+\fShiftVec-\gShiftVec\Bigr].
\end{align}
Variation of the potential with respect to $\sLp$ gives (see appendix
\ref{app:deriv}),
\begin{align}
\gLapse\partial e_{n}(S)/\partial\sLp & =\sEta\,Y_{n-1}\bigl(\fESp\gESp^{-1}\bigr)\,\fESp\,\Bigl[(\gLapse\gESp^{-1}+\fLapse\fESp^{-1})\,\sLv+\fShiftVec-\gShiftVec\Bigr]\nonumber \\
 & \qquad+\,\frac{1}{\sLt+1}\Bigl[\sLs^{\tr}\sEta\,Y_{n-1}(\fESp\gESp^{-1})\,\fESp\,(\gLapse\gESp^{-1}+\fLapse\fESp^{-1})\,\sLs^{-1}\Bigr]^{\tr}\sLv\nonumber \\
 & \qquad-\,\frac{1}{\sLt+1}\Bigl[\sEta\,Y_{n-1}\bigl(\fESp\gESp^{-1}\bigr)\,\fESp\,(\gLapse\gESp^{-1}+\fLapse\fESp^{-1})\Bigr]\,\sLv.\label{eq:NenS3}
\end{align}
Now, we need to address the residual spatial rotations $\sRs$ of
$\fESp$ (or $\gESp$) because, together with $\partial\mathcal{V}/\partial\sLp$,
we have to vary $\mathcal{V}$ with respect to $\sRs$. This can be
done by parametrizing $\sRs$ similarly to $\Lambda$, then varying
with respect to the parameters of $\sRs$. For instance, consider
the 2$\times$2 dimensional Euclidean metrics,\bSe\label{eq:2dmet}
\begin{alignat}{2}
\tilde{g} & =\begin{pmatrix}\tilde{N}^{2}+\tilde{e}^{2}\tilde{\nu}^{2} &  & \tilde{e}^{2}\tilde{\nu}\\
\tilde{e}^{2}\tilde{\nu} &  & \tilde{e}^{2}
\end{pmatrix}\!, & \qquad\tilde{E} & =\begin{pmatrix}\tilde{N} &  & 0\\
\tilde{e}\,\tilde{\nu} &  & \tilde{e}
\end{pmatrix}\!,\\
\tilde{f} & =\begin{pmatrix}\tilde{M}^{2}+\tilde{m}^{2}\tilde{\mu}^{2} &  & \tilde{m}^{2}\tilde{\mu}\\
\tilde{m}^{2}\tilde{\mu} &  & \tilde{m}^{2}
\end{pmatrix}\!, & \qquad\tilde{L}_{0} & =\begin{pmatrix}\tilde{M} &  & 0\\
\tilde{m}\tilde{\mu} &  & \tilde{m}
\end{pmatrix}\!.
\end{alignat}
\eSe Let $\tilde{R}$ be the rotation of the triangular zweibein
$\tilde{L}_{0}$ such that $\tilde{L}=\tilde{R}\tilde{L}_{0}$, where
from (\ref{eq:Xdec}), 
\begin{equation}
\tilde{R}(\varrho)=\begin{pmatrix}\sqrt{1-\varrho^{2}} &  & -\varrho\\
\varrho &  & \sqrt{1-\varrho^{2}}
\end{pmatrix}\!.
\end{equation}
The variation of $e_{1}(\tilde{L}\tilde{E}^{-1})$ with respect to
$\varrho$ gives the equation of motion,
\begin{equation}
\tilde{\nu}-\tilde{\mu}=(\tilde{M}\tilde{m}^{-1}+N\tilde{e}^{-1})r,\qquad r=\sqrt{1-\varrho^{2}}/\varrho\in\mathbb{R}.\label{eq:rshift}
\end{equation}
This is a ``shift difference'' condition which is similar to (\ref{eq:symca})
below. The solution to (\ref{eq:rshift}) symmetrizes the zweibeins
from (\ref{eq:2dmet}), which gives the square root $(\tilde{g}^{-1}\tilde{f})^{1/2}=\tilde{R}\tilde{L}_{0}\tilde{E}^{-1}$.
A similar derivation, this time for the full rotations, yields the
shift difference conditions for the symmetrization of the spatial
part $\gESp^{\tr}\sEta\sLs\fESp$.

The symmetric $\gESp^{\tr}\sEta\sLs\fESp$ implies the symmetric $\sLs^{\tr}\sEta Y_{n}(\fESp\gESp^{-1})$,
so (\ref{eq:NenS3}) becomes,
\begin{equation}
\gLapse\partial e_{n}(S)/\partial\sLv=\sEta\,Y_{n-1}(\fESp\gESp^{-1})\,\fESp\,\left[(\gLapse\gESp^{-1}+\fLapse\fESp^{-1})\sLv+\fShiftVec-\gShiftVec\right].
\end{equation}
Hence, the equations of motion with respect to both $\sLp$ and $\sRs$
are solved by,
\begin{equation}
\delShift\coloneqq(\gLapse\gESp^{-1}+\fLapse\fESp^{-1})\sLv+\fShiftVec-\gShiftVec=0,
\end{equation}
together with $\sEta\sLs\fESp\gESp^{-1}=(\sEta\sLs\fESp\gESp^{-1})^{\tr}$.
These two conditions are equivalent to the symmetrization condition,
\begin{equation}
\gE^{\tr}\eta\fE=(\gE^{\tr}\eta\fE)^{\tr},
\end{equation}
which gives the real square root $S=(\gMet^{-1}\fMet)^{1/2}$. Namely,
a congruence $S$ is self-adjoint $\gMet S=(\gMet S)^{\tr}$ (it is
a square root), if and only if \cite{Hassan:2014gta,Kocic:2018ddp},\bSe\label{eq:symc}
\begin{align}
\gShiftVec-\fShiftVec & =(\gLapse\gESp^{-1}+\fLapse\fESp^{-1})\sLv,\label{eq:symca}\\
\gESp^{\tr}\sEta\sLs\fESp & =(\gESp^{\tr}\sEta\sLs\fESp)^{\tr}.\label{eq:symcb}
\end{align}
\eSe In terms of the variables $D$ and $Q$ from \cite{Hassan:2011tf,Hassan:2011zd},
the spatial symmetrization (\ref{eq:symcb}) reads $\fSp D=(\fSp D)^{\tr}$
where $DQD=\gSp^{-1}\fSp$, $DQ=\sBs$, and $Q=\fESp^{-1}\sLs^{2}\fESp$.

Finally, the coupled system (\ref{eq:symc}) is equivalent to the
condition on metrics to have intersecting null cones with a common
timelike direction and a common spacelike hypersurface element \cite{Hassan:2017ugh}.
The first equation controls the separation between the null cones,
and the second ensures that the spatial shapes of the null cones properly
intersect.

\section{Discussion}

\label{sec:discussion}

The analysis in \cite{Hassan:2011tf,Hassan:2011zd} starts from the
square root in the bimetric potential, which gives the required redefinition
of the shift variable that is essential for the ghost-free proof in
the $\NPOne$ formalism. We have shown that this shift redefinition
and the spatial symmetrization come out from the equation of motion
for a general congruence between the metrics. The primary square root
naturally emerges as the on-shell condition in the covariant form,
which further clarifies the square root branch and type selection
in \cite{Hassan:2017ugh}.

In earlier versions of \cite{Hinterbichler:2012cn}, the authors proposed
a method for dealing with the local spatial rotation invariance using
constrained spatial vielbeins. This method is omitted in the most
recent version of the paper. The authors justify the removal in a
footnote, pointing out that it is not clear that such a method works
since solving the constraint  may introduce dependence on the lapse
or shift into the vielbeins. As shown in sec.~\ref{sec:np1}, variation
with respect to both $\sLp$ and the spatial rotation resolves this
issue. In fact, it produces the square root $S=(\gMet^{-1}\fMet)^{1/2}$,
which is compatible with the result from appendix C in \cite{Hinterbichler:2012cn}.

Multiple spin-2 fields interactions in \cite{Hinterbichler:2012cn}
and \cite{Hassan:2018mcw} rely on the vielbein formulation. In this
work, we only treat bimetric interactions. An extension to multimetric
congruences is possible using the generalized symmetric polynomials
introduced in~\cite{Hinterbichler:2012cn}. A similar remark holds
for interactions of multiple spin-2 fields beyond pairwise couplings~\cite{Hassan:2018mcw}.
For a singled out metric $g_{I}$, the interaction given in \cite{Hassan:2018mcw}
can be formulated as $\sqrt{-g_{I}}\det\!\big(\sum_{J=1}^{\mathcal{N}}\beta^{J}S_{IJ}\big)$
in terms of the congruences $g_{J}=S_{IJ}^{\tr}g_{I}S_{IJ}$ (note
the transitivity $S_{IJ}=S_{IK}S_{KJ}$). 

Even though the vielbein formulation seems more fundamental, when
it comes to partial differential equations and the causal propagation
of the matter fields, one still needs the metric inverses and contracted
covariant derivatives to write down the wave equation. This also holds
for the fermionic fields. In general, a well-posed system of partial
differential equations for arbitrary tensorial spacetimes carrying
predictive, interpretable, and quantizable matter requires a bi-hyperbolic
principal symbol~\cite{Raetzel:2010je,Schuller:2014jia,Schuller:2016onj}.
For bimetric theory, the principal symbol is the totally symmetrized
product $\gMet^{(\mu\nu}\fMet^{\rho\sigma)}$, which is bi-hyperbolic
if and only if the real square root of $\gMet^{\mu\rho}\fMet_{\rho\nu}$
exists. A similar property still lacks for the multivielbein/multimetric
formulation of spin-2 interactions.

\subsubsection*{Acknowledgments\vspace{-0.5ex}}

I am grateful to Fawad Hassan for the valuable discussions and suggestion
how to shorten the proof in section \ref{sec:cov}. I also thank Edvard
M\"{o}rtsell, Fawad Hassan, Francesco Torsello, and Marcus H\"{o}g\r{a}s
for the comments and careful reading of the draft.


\vfill
\emitAppendix

\section{Recovering indices}

\label{app:indices}

Below is the list of geometrical objects with their indices attached,

\begin{table}[H]
\noindent \centering{}\bgroup\renewcommand{\arraystretch}{1.1}%
\begin{tabular}{cccccccccc}
$\tud S{\mu}{\nu}$, &  & $\tdu{(S^{\tr})}{\nu}{\mu}$, &  & $\tud{(S^{-1})}{\nu}{\mu}$, & $\tdu{(S^{-1,\tr})}{\mu}{\nu}$, &  & $\sLP_{\mu\nu}$, &  & $\tud X{\mu}{\nu}$,\tabularnewline
$\gMet_{\mu\nu}$, &  & $\gSp_{ij}$, &  & $\gShiftVec^{i}$, & $\fMet_{\mu\nu}$, &  & $\fSp_{ij}$, &  & $\fShiftVec^{i}$,\tabularnewline
$\tud{\gE}A{\mu}$, &  & $\tdu{(\gE^{\tr})}{\mu}A$, &  & $\tud{(\gE^{-1})}{\mu}A$, & $\tud{\fE}A{\mu}$, &  & $\tdu{(\fE^{\tr})}{\mu}A$, &  & $\tud{(\fE^{-1})}{\mu}A$,\tabularnewline
$\tud{\gESp}ai$, &  & $\tdu{(\gESp^{\tr})}ia$, &  & $\tud{(\gESp^{-1})}ia$, & $\tud{\fESp}ai$, &  & $\tdu{(\fESp^{\tr})}ia$, &  & $\tud{(\fESp^{-1})}ia$,\tabularnewline
$\tud{\sB}{\mu}{\nu}$, &  & $\tud{\sBs}ij$, &  &  & $\tud{\fEs}ai$, &  & $\fEu^{i}$, &  & $\fEt_{i}$,\tabularnewline
$\eta_{AB}$, &  & $\sEta_{ab}$, &  & $\sLp^{a}$, & $(\sLp^{\tr})^{a}$, &  & $\sLv^{a}$, &  & $(\sLv^{\tr})^{a}$,\tabularnewline
$(\eta^{-1})^{AB}$, &  & $(\sEta^{-1})^{ab}$, &  & $\tud{\sI}ab$, & $\tud{\sLs}ab$. &  &  &  & \tabularnewline
\end{tabular}\egroup
\end{table}

\noindent For example:\enskip{} $\fEt^{\tr}\fEu\,\leftrightarrow\,\fEt_{i}\fEu^{i}$,\enskip{}
$\fEt\fEt^{\tr}\,\leftrightarrow\,\fEt_{i}\fEt_{j}$,\enskip{} $\sBs=\gESp^{-1}\fEs\,\leftrightarrow\,\tud{\sBs}ij=\tud{(\gESp^{-1})}ia\tud{\fEs}aj$,\enskip{}
and,
\begin{equation}
(\gLapse\gESp^{-1}+\fLapse\fESp^{-1})\,\sLv+\fShiftVec-\gShiftVec\quad\leftrightarrow\quad\Big[\gLapse\tud{(\gESp^{-1})}ia+\fLapse\tud{(\fESp^{-1})}ia\Big]\sLv^{a}+\fShiftVec^{i}-\gShiftVec^{i}.
\end{equation}
The adjoint of any $A$ with respect to $\gMet$ reads $(\tud{A^{\prime})}{\mu}{\nu}=(\gMet^{-1})^{\mu\rho}\tdu{(A^{\tr})}{\rho}{\sigma}\gMet_{\sigma\nu}$.
The functions $\Tr$ and $e_{n}$ contract the first with the last
index; hence, their argument must be a (1,1)-tensor.

\section{Detailed derivations}

\label{app:deriv}

\subsection*{Detailed derivations for section 2}

We start from (\ref{eq:Vpot}) where $S$ is not uniquely determined
by $\gMet,\fMet$: we have additional degrees of freedom in an orthogonal
transformation $X$ of $\gMet$, where $\gMet=X^{\tr}\gMet X$, such
that,
\begin{equation}
\fMet=S^{\tr}\gMet S=S_{0}^{\tr}X^{\tr}\gMet XS_{0},\quad S=XS_{0},
\end{equation}
where $S_{0}$ is a function only of $\gMet,\fMet$ which does not
depend on $X$. 

We parametrize $X$ using the Cayley transformation where,
\begin{equation}
X=(\gMet+\sLP)^{-1}(\gMet-\sLP),\quad X(\sLP)^{-1}=X(-\sLP),\quad X^{\tr}=(\gMet+\sLP)(\gMet-\sLP)^{-1},
\end{equation}
together with,
\begin{equation}
2(\gMet+\sLP)^{-1}\gMet=I+X,\qquad2(\gMet+\sLP)^{-1}\sLP=I-X.
\end{equation}
The variation of $X$ in terms of $\sLP$ is,
\begin{gather}
(\gMet+\sLP)X=\gMet-\sLP,\\
(\delta\sLP)\,X+(\gMet+\sLP)\delta X=-\delta\sLP,\\
\delta X=-(\gMet+\sLP)^{-1}\delta\sLP\,(I+X)=-2(\gMet+\sLP)^{-1}\delta\sLP\,(\gMet+\sLP)^{-1}\gMet.
\end{gather}
On the other hand, the variation of potential in terms of $X$ is,
\begin{align}
\delta V(S) & =\delta\left(\sum_{n=0}^{d}\beta_{n}e_{n}(S)\right)=\delta\left(\sum_{n=0}^{d}\frac{\beta_{n}}{n}\sum_{k=1}^{n}(-1)^{k}e_{n-k}(S)\,\Tr(S^{k})\right)\\
 & =\sum_{n=1}^{d}\beta_{n}\sum_{k=1}^{n}(-1)^{k}e_{n-k}(S)\,\Tr(S^{k-1}\delta S)\\
 & =\sum_{n=1}^{d}\beta_{n}\sum_{k=1}^{n}(-1)^{k}e_{n-k}(S)\,\Tr(X^{-1}S^{k}\delta X),
\end{align}
where in the last two steps we used $\delta S=\delta(XS_{0})=(\delta X)S_{0}=\delta X\,X^{-1}XS_{0}=\delta X\,X^{-1}S$
together with the cyclic property of the trace. 

After substituting $X^{\tr}=$ and $\delta X$ into $\delta V(S)$,
we get,
\begin{align}
\delta V(S) & =-2\sum_{n=1}^{d}\beta_{n}\sum_{k=1}^{n}(-1)^{k}e_{n-k}(S)\,\Tr(S^{k}(\gMet+\sLP)^{-1}\delta\sLP\,(\gMet+\sLP)^{-1}\gMet X^{-1})\\
 & =-2\sum_{n=1}^{d}\beta_{n}\sum_{k=1}^{n}(-1)^{k}e_{n-k}(S)\,\Tr(S^{k}(\gMet+\sLP)^{-1}\delta\sLP\,(\gMet+\sLP)^{-1}X^{\tr}\gMet)\\
 & =-2\sum_{n=1}^{d}\beta_{n}\sum_{k=1}^{n}(-1)^{k}e_{n-k}(S)\,\Tr(S^{k}(\gMet+\sLP)^{-1}\delta\sLP\,(\gMet-\sLP)^{-1}\gMet)\\
 & =-2\sum_{n=1}^{d}\beta_{n}\sum_{k=1}^{n}(-1)^{k}e_{n-k}(S)\,\Tr((\gMet-\sLP)^{-1}\,\gMet S^{k}(\gMet+\sLP)^{-1}\delta\sLP).
\end{align}
Since $p$ is skew-symmetric, we have $\delta\sLP^{\tr}=-\delta\sLP$
and, 
\begin{equation}
\delta\Tr(C\sLP)=\Tr(C\,\delta\sLP)=\Tr(\delta\sLP^{\tr}C^{\tr})=-\Tr(C^{\tr}\delta\sLP),
\end{equation}
that is \cite{Petersen:06mx},
\begin{equation}
\delta\Tr(C\sLP)=\frac{1}{2}\Tr\left[(C-C^{\tr})\delta\sLP\right],\quad\to\quad\partial\Tr(C\sLP)/\partial\sLP=\frac{1}{2}(C-C^{\tr}).
\end{equation}
Hence,
\begin{align}
\partial V(S)/\partial\sLP & =-\sum_{n=1}^{d}\beta_{n}\sum_{k=1}^{n}(-1)^{k}e_{n-k}(S)\times\nonumber \\
 & \qquad\times\left[\left((\gMet-\sLP)^{-1}\gMet S^{k}(\gMet+\sLP)^{-1}-\left((\gMet-\sLP)^{-1}\gMet S^{k}(\gMet+\sLP)^{-1}\right)^{\tr}\right)\right]\\
 & =-\sum_{n=1}^{d}\beta_{n}\sum_{k=1}^{n}(-1)^{k}e_{n-k}(S)\left[(\gMet-\sLP)^{-1}\left(\gMet S^{k}-(\gMet S^{k})^{\tr}\right)(\gMet+\sLP)^{-1}\right].
\end{align}
The equations of motion for $\sLP$ are $\partial V/\partial\sLP=0$,
i.e., having a nonsingular $\gMet$ (and $\gMet\pm\sLP$), 
\begin{align}
\sum_{n=1}^{d}\beta_{n}\sum_{k=1}^{n}(-1)^{k}e_{n-k}(S)\left[S^{k}-\left(\gMet^{-1}S^{\tr}\gMet\right)^{k}\right] & =0.
\end{align}

\subsection*{Detailed derivations for section 3, part 1}

Let us introduce the relation $\stackrel{e_{n}}{=\joinrel=}$ which
indicates that the two expressions are equal after $e_{n}$ is applied
to both sides of the equation. Using the cyclic property of $e_{n}$,
we obtain,
\begin{equation}
S\stackrel{e_{n}}{=\joinrel=}\frac{1}{N}\begin{pmatrix}\fEut+\fEt^{\tr}(\fEu-\gShiftVec)\\
\gESp^{-1}\fEs(\fEu-\gShiftVec)
\end{pmatrix}\!\begin{pmatrix}1\\
0
\end{pmatrix}^{\tr}+\begin{pmatrix}0 &  & \fEt^{\tr}\\
0 & \, & \gESp^{-1}\fEs
\end{pmatrix}=\frac{1}{\gLapse}UW^{\tr}+\sB,
\end{equation}
where,
\begin{equation}
U\coloneqq\begin{pmatrix}\fEut\\
0
\end{pmatrix}+\sB\begin{pmatrix}0\\
\fEu-\gShiftVec
\end{pmatrix}\!,\quad W\coloneqq\begin{pmatrix}1\\
0
\end{pmatrix}\!,\quad\sB\coloneqq\begin{pmatrix}0 &  & \fEt^{\tr}\\
0 &  & \sBs
\end{pmatrix}\!,\quad\sBs\coloneqq\gESp^{-1}\fEs.
\end{equation}
Then,
\begin{align}
S & \stackrel{e_{n}}{=\joinrel=}\left[\frac{1}{\gLapse}\begin{pmatrix}1\\
-\gShiftVec
\end{pmatrix}\!\begin{pmatrix}1\\
0
\end{pmatrix}^{\tr}+\begin{pmatrix}0 &  & 0\\
0 &  & \gESp^{-1}
\end{pmatrix}\right]\!\begin{pmatrix}\fEut+\fEt^{\tr}\fEu &  & \fEt^{\tr}\\
\fEs\fEu &  & \fEs
\end{pmatrix}\\
 & \stackrel{e_{n}}{=\joinrel=}\begin{pmatrix}0 &  & 0\\
0 &  & \gESp^{-1}
\end{pmatrix}\!\begin{pmatrix}\fEut+\fEt^{\tr}\fEu &  & \fEt^{\tr}\\
\fEs\fEu &  & \fEs
\end{pmatrix}\!\left[\frac{1}{\gLapse}\begin{pmatrix}1\\
-\gShiftVec
\end{pmatrix}\!\begin{pmatrix}1\\
0
\end{pmatrix}^{\tr}+\begin{pmatrix}0 &  & 0\\
0 &  & \sI
\end{pmatrix}\right]\\
 & =\begin{pmatrix}\fEut+\fEt^{\tr}\fEu &  & \fEt^{\tr}\\
\gESp^{-1}\fEs\fEu &  & \gESp^{-1}\fEs
\end{pmatrix}\!\left[\frac{1}{\gLapse}\begin{pmatrix}1\\
-\gShiftVec
\end{pmatrix}\begin{pmatrix}1\\
0
\end{pmatrix}^{\tr}+\begin{pmatrix}0 &  & 0\\
0 &  & \sI
\end{pmatrix}\right]\!.
\end{align}
Using the following identity (which holds for arbitrary $B$ and vectors
$u,w$),
\begin{eqnarray}
e_{n}(B+uw^{\tr}) & = & e_{n}(B)+\sum_{k=1}^{n}(-1)^{k-1}e_{n-k}(B)\,w^{\tr}B^{k-1}u,\label{eq:esp-2}
\end{eqnarray}
we immediately have,
\begin{equation}
e_{n}(S)=e_{n}(\sB)+\frac{1}{\gLapse}\sum_{k=1}^{n}(-1)^{k-1}e_{n-k}(\sB)\,W^{\tr}\sB^{k-1}U.
\end{equation}
Note that $e_{n}(\sB)=e_{n}(\sBs)$ and $W^{\tr}U=\fEut$; thus,
\begin{equation}
\gLapse\,e_{n}(S)=\gLapse\,e_{n}(\sBs)+\fEut e_{n-1}(\sBs)+\sum_{k=1}^{n}(-1)^{k-1}e_{n-k}(\sBs)\,\begin{pmatrix}1\\
0
\end{pmatrix}^{\tr}\sB^{k}\begin{pmatrix}0\\
\fEu-\gShiftVec
\end{pmatrix}\!.
\end{equation}
Then from,
\begin{equation}
\sB^{k}=\begin{pmatrix}0 &  & \fEt^{\tr}\sBs^{k-1}\\
0 &  & \sBs^{k}
\end{pmatrix}\!,
\end{equation}
it follows,
\begin{equation}
\gLapse\,e_{n}(S)=\gLapse\,e_{n}(\sBs)+\fEut\,e_{n-1}(\sBs)+\sum_{k=1}^{n}(-1)^{k-1}e_{n-k}(\sBs)\,\fEt^{\tr}\sBs^{k-1}(\fEu-\gShiftVec)\,.
\end{equation}
where $e_{n}(\sBs)=0$ for $n>d-1$. 

Assuming that we are in a frame where both $\fMet$ and $\gMet$ admit
proper decomposition (both $\gE$ and $\fE$ can be triangularized),
we get,
\begin{align}
\gLapse\,e_{n}(S) & =\gLapse\,e_{n}(\sBs)+\sLt^{-1}\fLapse\,e_{n-1}(\sBs)\,+\nonumber \\
 & \qquad+\,\sum_{k=1}^{n}(-1)^{k-1}e_{n-k}(\sBs)\,\fEt^{\tr}\sBs^{k-1}(\fLapse\fESp^{-1}\sLv+\fShiftVec-\gShiftVec).
\end{align}
Note that $\sBs=\gESp^{-1}\fEs$ corresponds to the congruence,
\begin{gather}
\sBs^{\tr}\gSp\sBs=\fEs^{\tr}\sEta\fEs=\fSp+\fEt\fEt^{\tr}=\fSp(\sI+\fSp^{-1}\fEt\fEt^{\tr})=\fSp Q=Q^{\tr}\fSp,\qquad Q\coloneqq\fESp^{-1}\sLs^{2}\fESp.
\end{gather}

\subsection*{Note on derivations for section 3, part 2}

For the calculation of (\ref{eq:NenS2}), see the derivation tracks
6 and 7 in the ancillary Mathematica notebook. The variation $\partial\mathcal{V}/\partial\sLp$
in (\ref{eq:NenS3}) is done using the following identities (for more
details see Part 3 in the Mathematica notebook),
\begin{gather}
\partial(\sLp^{\tr}A)/\partial\sLp=A^{\tr},\\
\shortintertext{and,}\partial(A\frac{\sLp}{\sLt+1})/\partial\sLp=\frac{1}{\sLt+1}A\sLs^{-1},
\end{gather}
where $A$ is an arbitrary matrix that does not depend on $\sLp$
and,
\begin{equation}
\sLs^{-1}=\sI-\frac{1}{\sLt(\sLt+1)}\sLp\sLp^{\tr}\sEta.
\end{equation}

\section{Bimetric actions with a scalar potential}

\label{app:bimpot}

Here we highlight GR-type actions where the interaction between metrics
is given through a scalar potential. Consider the metric fields $\gMet$
and $\fMet$ where the dynamics of each metric is governed by the
Einstein\textendash Hilbert term. Each diffeomorphism group $\op{Diff}(\gMet)$
and $\op{Diff}(\fMet)$ acts separately on its own metric. In the
interacting case where the interaction is given through a scalar potential
$\mathcal{V}(\gMet,\fMet)$, the symmetry of the full action $(\phi_{1},\phi_{2})\in\op{Diff}(\gMet)\times\op{Diff}(\fMet)$
must be reduced to the diagonal group of common diffeomorphisms where
$\phi_{1}=\phi_{2}$ by the theorem in \cite{Boulanger:2000bp,Boulanger:2000rq}.
This demands that the interaction term depends only on the scalars
one can make with two metrics \cite{Damour:2002ws}. More precisely,
the common diffeomorphism invariance restricts the scalar potential
to depend only on the invariants of the (1,1) tensor field $\gMet^{-1}\tilde{\fMet}$
where $\varphi$ is an overall diffeomorphism $\phi_{1}=\varphi\circ\phi_{2}$
and $\tilde{\fMet}=\varphi^{*}\fMet$ is the pullback of $\fMet$
by $\varphi$,
\begin{equation}
\tilde{\fMet}_{\mu\nu}=\partial_{\mu}\varphi^{\bar{\alpha}}\partial_{\nu}\varphi^{\bar{\beta}}\fMet_{\bar{\alpha}\bar{\beta}}.
\end{equation}
The map $\varphi$ is part of the local trivialization of the tangent
bundles. It gives rise to St\"uckelberg fields $\varphi^{\bar{\alpha}}$
that do not introduce new dynamics into the theory; hence, $\varphi$
can be fixed to be the identity map in the unitary gauge where $\partial_{\mu}\varphi^{\bar{\alpha}}=\delta_{\mu}^{\bar{\alpha}}$,
sloppily setting $\tilde{\fMet}=\fMet$.

The analysis in section \ref{sec:cov} would not change if we had
$\tilde{\fMet}=\tilde{S}^{\tr}\gMet\tilde{S}$ and $\varphi^{*}\tilde{\fMet}=S^{\tr}\gMet S$,
\begin{equation}
\gMet^{-1}S^{\tr}\gMet=\gMet^{-1}S^{\tr}\gMet\,SS^{-1}=\gMet^{-1}\varphi^{*}\tilde{\fMet}\,S^{-1}=\gMet^{-1}\fMet\,S^{-1}=AS^{-1},
\end{equation}
since $\varphi^{*}\tilde{\fMet}=\Phi^{\tr}\tilde{S}^{\tr}\gMet\tilde{S}\Phi$
with $S=\tilde{S}\Phi$, that is, $\tud S{\mu}{\nu}=\tud{\tilde{S}}{\mu}{\bar{\alpha}}\partial_{\nu}\varphi^{\bar{\alpha}}$
where $\tud{\Phi}{\bar{\alpha}}{\mu}=\partial_{\mu}\varphi^{\bar{\alpha}}$,
and,
\begin{equation}
A=\gMet^{-1}\fMet=\gMet^{-1}\Phi^{\tr}\tilde{\fMet}\Phi.
\end{equation}

\paragraph*{St\"uckelberg trick.}

A congruence in the most general form reads,
\begin{equation}
\fMet_{\bar{\alpha}\bar{\beta}}=\tud S{\mu}{\bar{\alpha}}\tud S{\nu}{\bar{\beta}}\gMet_{\mu\nu}.
\end{equation}
where $\tud S{\mu}{\bar{\alpha}}=\partial_{\bar{\alpha}}\varphi^{\gamma}\tud{\tud X{\mu}{\rho}S_{0}}{\rho}{\gamma}$
and $\varphi$ is a diffeomorphism such that $\partial_{\bar{\alpha}}\varphi^{\gamma}$
is the pullback (differential map) that is ``moving'' $\fMet$ to
the same tangent bundle where $\gMet$ lives ($\tilde{\fMet}=\varphi^{*}\fMet$).
To be able to write down $e_{n}(S)$, we must use $\tud S{\mu}{\nu}=\tud S{\mu}{\bar{\alpha}}\partial_{\nu}\varphi^{\bar{\alpha}}$.
The redundant gauge degrees of freedom in $\varphi$ do not introduce
new dynamics into the theory. The Einstein\textendash Hlibert term
is ``blind'' to both $\partial_{\bar{\alpha}}\varphi^{\gamma}$
and $\tud X{\mu}{\rho}$. Also,
\begin{equation}
\frac{\delta}{\delta\varphi^{\bar{\alpha}}}\mathcal{V}(\gMet,\fMet)=0.
\end{equation}
This equation will not give rise to any new dynamics since it is implied
by the Bianchi constraint (for instance, see \cite{Schmidt-May:2015vnx}),
\begin{equation}
\nabla^{\mu}V_{\mu\nu}(\gMet,\fMet)=0,\qquad\text{where }\,V_{\mu\nu}(\gMet,\fMet)\coloneqq\frac{-2}{\sqrt{-\gMet}}\frac{\delta}{\delta\gMet^{\mu\nu}}\left[\sqrt{-\gMet}\,\mathcal{V}(\gMet,\fMet)\right].
\end{equation}
Indeed, a gauge transformation under the diagonal group of diffeomorphisms
gives,
\begin{equation}
\nabla^{\mu}V_{\mu\nu}(\gMet,\fMet)=\frac{\delta\mathcal{V}}{\delta\varphi^{\bar{\alpha}}}\partial_{\nu}\varphi^{\bar{\alpha}},
\end{equation}
where $\partial_{\nu}\varphi^{\bar{\alpha}}$ is nonsingular.


\clearpage
\ifprstyle

\bibliographystyle{apsrev4-1}
\bibliography{bim-cong}

\else

\bibliographystyle{JHEP}
\bibliography{bim-cong}

\fi
\end{document}